\documentclass[prb,amsmath,showpacs,twocolumn,floatfix]{revtex4-1}

\usepackage{amssymb}
\usepackage{graphicx}
\usepackage{subfigure}
\usepackage{color}
\usepackage{multirow}
\usepackage{verbatim}
\usepackage{array}

\begin{document}


\title{
Multiferroic Bi$_2$NiMnO$_6$ Thin Films: A Computational Prediction
}

\author{Oswaldo Di\'eguez$^{1,2}$ and Jorge \'I\~niguez$^{3}$}

\affiliation{\vspace{0.3cm}
             $^1$Department of Materials Science and Engineering, 
             Faculty of Engineering, Tel Aviv University,
             Tel Aviv 69978, Israel 
             \vspace{0.3cm} \\
             $^2$The Raymond and Beverly Sackler Center for Computational
             Molecular and Materials Science,
             Tel Aviv University,
             Tel Aviv 69978, Israel
             \vspace{0.3cm} \\
             $^3$Materials Research and Technology Department, Luxembourg
             Institute of Science and Technology, 5 avenue des Hauts-Fourneaux,
             L-4362 Esch/Alzette, Luxembourg 
             \vspace{0.3cm} \\}


\begin{abstract}
We report first-principles calculations for one of the few
materials that is believed to be a ferroelectric ferromagnet, Bi$_2$NiMnO$_6$.
Our calculations show that, contrary to what it has been reported so far, 
bulk Bi$_2$NiMnO$_6$ does not have a polarization. Instead, like BiMnO$_3$, 
it crystallizes into
a centrosymmetric structure with space group $C2/c$. 
We also predict that Bi$_2$NiMnO$_6$ will indeed be a ferroelectric ferromagnet
if it is grown as an epitaxial film on a substrate with in-plane square
symmetry and a lattice constant around 4~\AA, such as BaTiO$_3$ or 
PbZr$_{1-x}$Ti$_{x}$O$_{3}$.
\end{abstract}

\date{\today}





\maketitle


\section{Introduction}

The possibility of manipulating the magnetization of a material by using an
electric field is probably the most attractive envisioned application 
of magnetoelectric multiferroics---ferroelectrics with
magnetic ordering.\cite{Bibes2008NM}
Even if the mechanisms responsible for ferroelectricity and magnetism are
somewhat exclusive of each other,\cite{Hill2000JPCB, Bhattacharjee2009PRL}
in the last decade a large research effort has gone into searching for 
these materials.\cite{Fiebig2005JPAP, Prellier2005JPCM,
Khomskii2006JMMM, Eerenstein2006N, Rao2007JMC, Ramesh2007NM, Tokura2007JMMM,
Cheong2007NM, Nan2008JAP, Wang2009AP, Akbashev2011RCR, Ma2011AM,
Pyatakov2012PU}
This effort has mainly focused on two groups of complex
oxides: those where different species are responsible for the polarization
and the magnetism, and those where the magnetic ordering breaks the inversion
simmetry of the structure to create a small polarization.
BiFeO$_3$ belongs to the first group and it is by far the most studied 
multiferroic,\cite{Catalan2009AM} mainly because it keeps both its ferroic
orderings
well above room temperature; it is also relatively easy to prepare in bulk
and film form, and it has a simple crystal structure---a perovskite where 
inversion symmetry is broken to accommodate the lone pair of Bi in the $A$ site,
while the $B$ site harbors the Fe ions whose $d$ electrons are responsible for
magnetism.
However, the ferromagnetic component in BiFeO$_3$ is tiny;
instead, the spins of two neighboring Fe ions are almost perfectly
antiparallel.
In the difficult search for single-phase ferroelectric ferromagnets that would
allow for a direct hysteresis loop of magnetization with electric field
some candidate materials have been proposed. 
Examples include EuTiO$_3$ (although ferromagnetism only settles at
around 4 K\cite{Lee2010N}), LuFe$_2$O$_4$ (although whether this is a
ferroelectric is still under debate\cite{Angst2013PSS,Yang2015APL}), 
Fe$_3$O$_4$ (although the exact structure that arises below the Verwey 
transitions is not yet understood\cite{vanderBrink2008JPCM}), CoCr$_2$O$_4$
(although both the magnetization and the polarization are very
small\cite{Yamasaki2006PRL}), and,
more recently, the metastable $\epsilon$-Fe$_2$O$_3.$\cite{Gich2014AM}

Researchers have also explored perovskite oxides similar to BiFeO$_3$, but
with other transition-metal ions instead of Fe.
BiMnO$_3$ is the only member of this group that displays strong
ferromagnetism. Initial reports attributed a polar $C2$ space group
to bulk BiMnO$_3$,\cite{Atou1999JSSC,Moreira2002PRB} but
more recent studies agree in that
this is a paraelectric with $C2/c$ symmetry.\cite{Belik2007JACS,
Montanari2007PRB, Baettig2007JACS}
It is possible to change the structure of this material by growing it
as a epitaxial thin film, although the calculations of Spaldin and
Hatt\cite{Hatt2009EPJB}
showed that when the substrate-imposed distortion is small enough to keep the 
ferromagnetism in BiMnO$_3$, then a polarization does not develop, and our 
calculations\cite{Dieguez2015PRB}
showed that when the distortion is large enough to create a 
large polarization, then ferromagnetism turns into
antiferromagnetism.
Another way to try to modify the properties of these oxides is to
add a second species in one of the sites of the perovskite.
Azuma and coworkers reasoned that the Goodenough-Kanamori 
rules\cite{Goodenough1958JPCS,Kanamori1959JPCS} predict
a ferromagnet if Mn and Ni shared the sites inside O$_6$ octahedra in a 
rock-salt pattern;
when they prepared Bi$_2$NiMnO$_6$ by high-pressure synthesis, they indeed
measured large parallel magnetic moments,\cite{Azuma2005JACS} which 
persisted up to a Curie temperature of 140 K.
After their synchrotron X-ray powder diffraction, they concluded that the 
material shows a heavily distorted double perovskite structure where the 
Ni$^{2+}$ and Mn$^{4+}$ ions are indeed ordered in a rock-salt configuration;
they assigned the space group $C2$ to this crystal.
Later, first-principles calculations characterized further this structure and
quoted a value of the polarization around 20
$\mu$C/cm$^2$.\cite{Ciucivara2007PRB, Zhao2012AIPA, Zhao2012AIPAb}

Unlike those previous studies, our
first-principles calculations 
show that bulk Bi$_2$NiMnO$_6$ is actually
a paraelectric material with $C2/c$ space group,
the same situation as with BiMnO$_3$. 
However, we predict that when Bi$_2$NiMnO$_6$ films are
grown under achievable tensile epitaxial strain, it will
indeed become a ferroelectric ferromagnet with a large polarization 
(70 $\mu$C/cm$^2$) and a magnetization above 2~$\mu_{\rm B}$ per transition
metal cation, as in the
bulk compound.\cite{Azuma2005JACS}
We describe the methodology we have used for our calculations in Section II,
present our results for the bulk material in Section III.A and for 
epitaxial films in Section III.B, and summarize the implications of our work
in Section IV.


\section{Methods}

Our first-principles calculations are based on density-functional theory
(DFT).\cite{Hohenberg1964PR,Kohn1965PR}
Following our previous study of
BiMnO$_3$,\cite{Dieguez2015PRB} we used two methods
to treat the localized $d$ orbitals of Ni$^{2+}$ and Mn$^{4+}$:
(i) DFT with a ``Hubbard $U$'',\cite{Liechtenstein1995PRB} using 
$U_{\rm Ni} = 1$~eV, $J_{\rm Ni} = 0$~eV, $U_{\rm Mn} = 4$~eV, and
$J_{\rm Mn} = 1$~eV;
and (ii) DFT with the HSE06 hybrid functional.\cite{Krakau2006JCP}
For all our calculations we have used the {\sc Vasp}\cite{VASP} code;
for this system, the second method demands two orders of magnitude more
computer time than 
the first, but it predicts band gaps for solids that are much closer to
experimental results,\cite{Krakau2006JCP} and in particular it performs well
for perovskite oxides such as BiFeO$_3$.\cite{Stroppa2010PCCP}

We used the Perdew-Burke-Ernzerhof DFT exchange-correlation functional
adapted to solids (PBEsol).\cite{Perdew2008PRL}
To treat the ionic cores we resorted to the projector augmented-wave
method,\cite{Blochl1994PRB} solving for the following electrons:
Ni's and Mn's $3p$, $3d$, and $4s$;
Bi's $5d$, $6s$, and $6p$;
and O's $2s$ and $2p$.
The plane-wave basis set kinetic-energy cutoff was 500 eV.
We performed integrations in the Brillouin zone using $k$-point grids
with densities similar to that of the
$6 \times 6 \times 6$ mesh for a 5-atom perovskite
unit cell.


\section{Results}

\subsection{Bulk Phases}

\begin{figure}
\centering
\subfigure[]{
\includegraphics[width=30mm, angle=90]{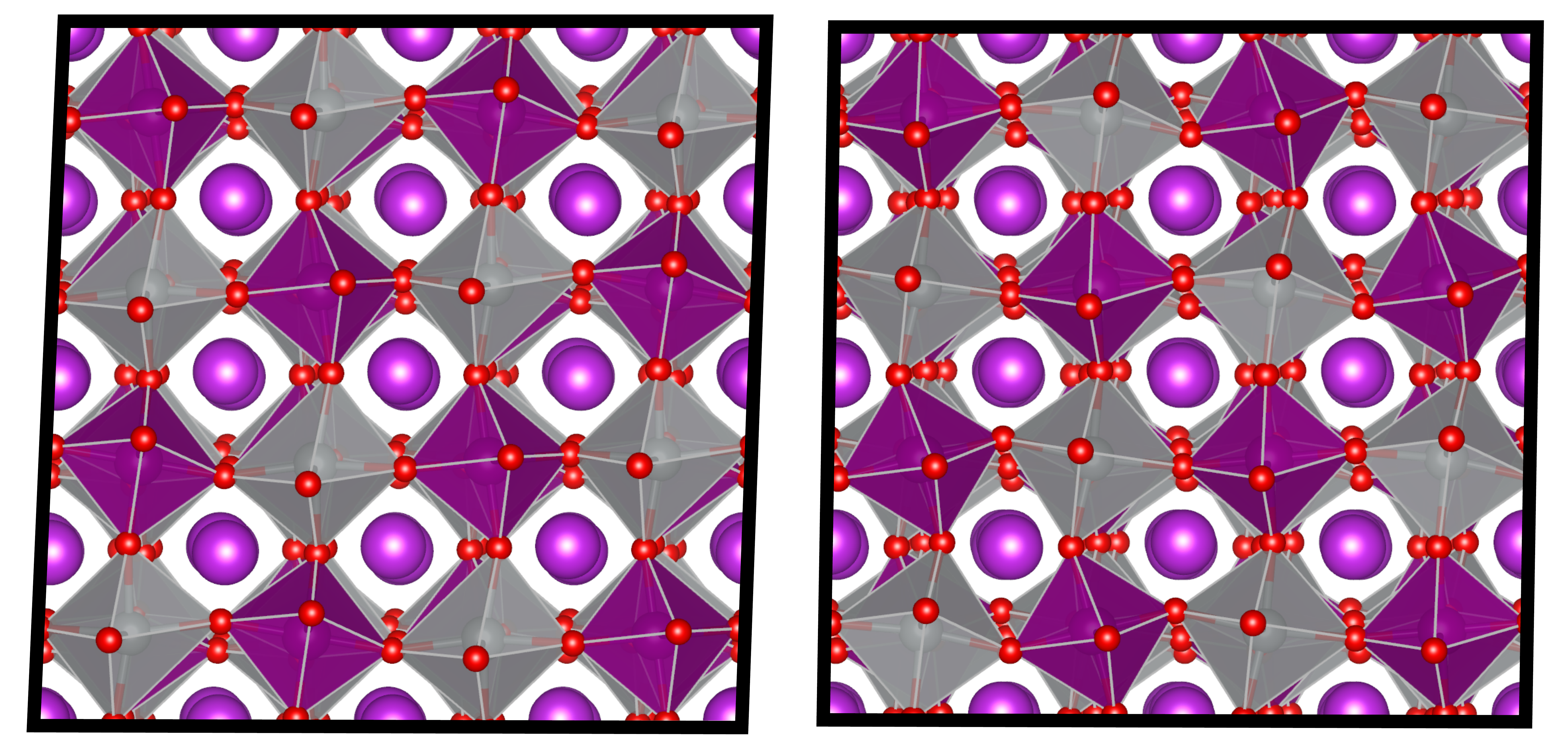}
}
\\
\subfigure[]{
\includegraphics[width=30mm, angle=90]{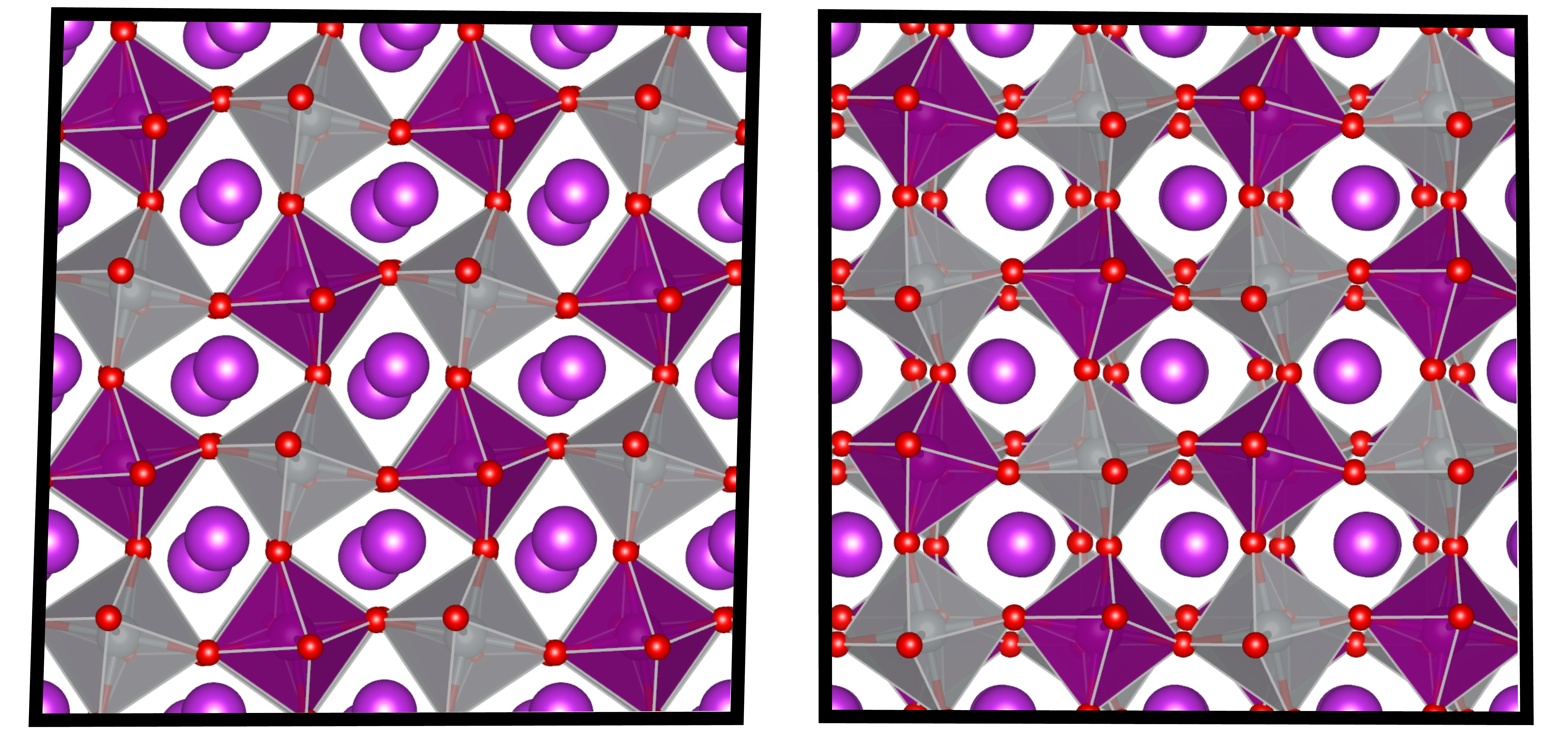}
}
\\
\subfigure[]{
\includegraphics[width=30mm, angle=90]{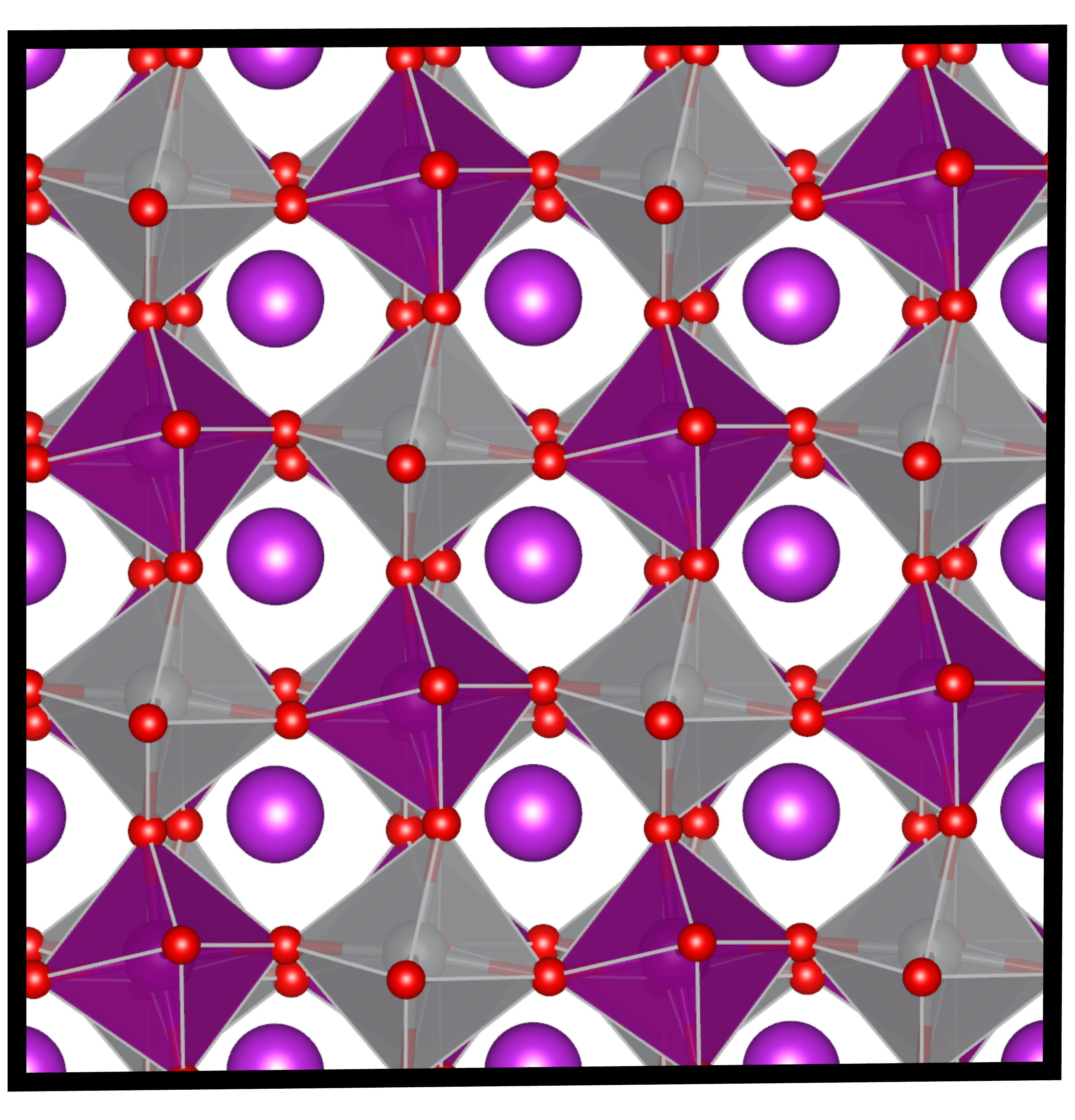}
}
\caption{(Color online.)
Inequivalent 
side views of a pseudocubic unit cell for bulk Bi$_2$NiMnO$_6$ structures that
correspond to energy minima according to
our DFT+$U$ calculations: (a) $C2/c$, 
(b) $P2_1/n$, and (c) $R3$.
}
\label{fig_structures}
\end{figure}

Our first set of calculations involves the optimization of Bi$_2$NiMnO$_6$
bulk structures that might be competitive in energy with the 
ground state.
As mentioned in the Introduction, previous experimental and computational 
studies consider that this ground state belongs to the $C2$ space group;
Refs.~\onlinecite{Azuma2005JACS} and \onlinecite{Ciucivara2007PRB} report
the lattice parameters and Wickoff positions of the atoms in the crystal 
unit cell, and we have run optimizations starting from those configurations.
In the future we call this structure GS---the structure that the bulk material
displays at low temperature and low pressure.
Another relevant phase is the one observed at high temperature with
space group $P2_1/n$, fully described in Ref.~\onlinecite{Takata2005JJSPPM};
this phase is analogous to the $Pnma$ phase that appears in many
Bi$M$O$_3$ perovskites at high pressure and/or high
temperature,\cite{Belik2012JSSC} but with reduced symmetry because of the
superimposed rock-salt pattern of Ni$^{+2}$ and Mn$^{+4}$ cations.
By analogy with our previous paper about BiMnO$_3$,\cite{Dieguez2015PRB}
we call this paraelectric phase $p$.
Based on our previous experience in the search of new phases of 
BiFeO$_3$,\cite{Dieguez2011PRB} BiCoO$_3$,\cite{Dieguez2011PRL} and 
BiMnO$_3$,\cite{Dieguez2015PRB} we also relaxed ferroelectric
phases similar to the rhombohedral ground state of BiFeO$_3$ (so-called $R$
phases) and to the supertetragonal ground state of BiCoO$_3$
(so-called $T$ phases); we have enforced the same rock-salt cation pattern
known to exist in both experimentally characterized phases of Bi$_2$NiMnO$_6$.
We have optimized these structures using the DFT+$U$ and the HSE06 
methods, obtaining similar results with both approaches regarding their 
structural details.
The atomic structure of the resulting optimized structures (with forces
converged below 0.015 eV/\AA~and stress components below 0.1 GPa) is
shown in Figure \ref{fig_structures}, while
Table~\ref{tab_bulk} contains the values of several magnitudes of interest
for these phases.

\begin{table*}
\setlength{\extrarowheight}{0.5mm}
\setlength{\tabcolsep}{5pt}
\caption{
Properties of Bi$_2$NiMnO$_6$ phases that are local energy minima 
according to our calculations (with DFT+$U$ and HSE06), and 
comparison with experiment (from Refs.~\onlinecite{Azuma2005JACS}
and \onlinecite{Takata2005JJSPPM}).
We report the space group, lattice parameters,
lattice angles, Wickoff positions, polarization $P$, and 
energy difference with the GS phase $\Delta E$.
GS, $p$, and $R$ label the ground-state phase, the high-temperature
paraelectric phase, and the $R$ rhombohedral phase found in this study, 
respectively.
}
\begin{tabular}{rrccc}
\hline\hline
 Phase & Properties
&  DFT+$U$  &  HSE06  &  Exp. \\  
\hline
GS            &  Space group            &  $C2/c$   &  $C2/c$   &  $C2$      \\
              &  $a$ (\AA)              &  9.3871   &  9.3523   &  9.4646    \\
              &  $b$ (\AA)              &  5.3739   &  5.3558   &  5.4230    \\
              &  $c$ (\AA)              &  9.5355   &  9.4679   &  9.5431    \\
              &  $\beta$ ($^\circ$)     &  107.64   &  107.66   &  107.82    \\
              &                    Mn 1 & $( 0.2500, 0.2500, 0.0000 )$
                                        & $( 0.2500, 0.2500, 0.0000 )$
                                        & $( 0.257 , 0.250 , 0.001  )$       \\
              &                    Ni 1 & $( 0.0000, 0.2648, 0.2500 )$
                                        & $( 0.0000, 0.2609, 0.2500 )$
                                        & $( 0.000 , 0.252 , 0.250  )$       \\
              &                    Ni 2 & --                            
                                        & --                           
                                        & $( 0.000 , 0.737 , 0.750  )$       \\
              &                    Bi 1 & $( 0.6308, 0.2323, 0.1235 )$
                                        & $( 0.6314, 0.2230, 0.1248 )$
                                        & $( 0.633 , 0.214 , 0.128  )$       \\
              &                    Bi 2 & --                           
                                        & --                           
                                        & $( 0.631 , 0.772 , 0.627  )$       \\
              &                     O 1 & $( 0.5897, 0.1970, 0.5833 )$
                                        & $( 0.5914, 0.1949, 0.5823 )$
                                        & $( 0.611 , 0.176 , 0.599  )$       \\
              &                     O 2 & $( 0.1582, 0.0385, 0.3860 )$
                                        & $( 0.1602, 0.0344, 0.3860 )$
                                        & $( 0.146 , 0.013 , 0.386  )$       \\
              &                     O 3 & $( 0.3498, 0.0350, 0.1577 )$
                                        & $( 0.3493, 0.0374, 0.1565 )$
                                        & $( 0.333 ,-0.021 , 0.163  )$       \\
              &                     O 4 & --                           
                                        & --                           
                                        & $( 0.920 , 0.279 , 0.430  )$       \\
              &                     O 5 & --                           
                                        & --                           
                                        & $( 0.377 , 0.941 , 0.649  )$       \\
              &                     O 6 & --                           
                                        & --                           
                                        & $( 0.662 , 0.453 , 0.876  )$       \\
              &  $P$ ($\mu$C/cm$^2$)    &  0        &  0        &  N/A       \\
              &  $\Delta E$ (meV/f.u.)  &  0        &  0        &  --        \\
\hline
$p$           &  Space group            &  $P2_1/n$ &  $P2_1/n$ &  $P2_1/n$  \\
              &  $a$ (\AA)              &  5.3590   &  5.3182   &  5.4039    \\
              &  $b$ (\AA)              &  5.5522   &  5.5262   &  5.5668    \\
              &  $c$ (\AA)              &  7.6517   &  7.6007   &  7.7330    \\
              &  $\beta$ ($^\circ$)     &  90.00    &  90.00    &  90.166    \\
              &                    Mn 1 & $( 0.0000, 0.5000, 0.0000 )$
                                        & $( 0.0000, 0.5000, 0.0000 )$
                                        & $( 0.0000, 0.5000, 0.0000 )$       \\
              &                    Ni 1 & $( 0.5000, 0.0000, 0.0000 )$
                                        & $( 0.5000, 0.0000, 0.0000 )$
                                        & $( 0.5000, 0.0000, 0.0000 )$       \\
              &                    Bi 1 & $( 0.0086, 0.0549, 0.2509 )$
                                        & $( 0.0072, 0.0553, 0.2512 )$
                                        & $( 0.0049, 0.0468, 0.2510 )$       \\
              &                     O 1 & $( 0.3067, 0.2878, 0.4589 )$
                                        & $( 0.3109, 0.2860, 0.4589 )$
                                        & $( 0.280 , 0.279 , 0.477  )$       \\
              &                     O 2 & $( 0.2921, 0.3040, 0.0386 )$ 
                                        & $( 0.2907, 0.3080, 0.0380 )$
                                        & $( 0.281 , 0.281 , 0.053  )$       \\
              &                     O 3 & $( 0.58372 -0.0222, 0.2556 )$
                                        & $( 0.5851, -0.0238, 0.2583 )$
                                        & $( 0.594 , -0.022,  0.252  )$       \\
              &  $P$ ($\mu$C/cm$^2$)    &  0        &  0        &  0         \\
              &  $\Delta E$ (meV/f.u.)  & -30       &  24       &  --        \\
\hline
$R$           &  Space group            &  $R3$     &  $R3$     &  --        \\
              &  $a$ (\AA)              &  5.4526   &  5.4428   &  --        \\
              &  $\alpha$ ($^\circ$)    &  60.35    &  60.02    &  --        \\
              &                    Mn 1 & $( 0.7250, 0.7250, 0.7250 )$ 
                                        & $( 0.7208, 0.7208, 0.7208 )$
                                        &                                    \\
              &                    Ni 1 & $( 0.2284, 0.2284, 0.2284 )$
                                        & $( 0.2255, 0.2255, 0.2255 )$
                                        & --                                 \\
              &                    Bi 1 & $( 0.0000, 0.0000, 0.0000 )$
                                        & $( 0.0000, 0.0000, 0.0000 )$
                                        & --                                 \\
              &                    Bi 2 & $( 0.4985, 0.4985, 0.4985 )$
                                        & $( 0.4987, 0.4987, 0.4987 )$
                                        & --                                 \\
              &                     O 1 & $( 0.4114, -0.0566, 0.5483 )$
                                        & $( 0.4122, -0.0646, 0.5450 )$
                                        & --                                 \\
              &                     O 2 & $( 0.0330, 0.4609, -0.0967 )$
                                        & $( 0.0225, 0.4580, -0.1000 )$
                                        & --                                 \\
              &  $P$ ($\mu$C/cm$^2$)    &  70       &  79       &  --        \\
              &  $\Delta E$ (meV/f.u.)  & -4        &  17       &  --        \\
\hline\hline
\end{tabular}
\label{tab_bulk}
\end{table*}

The main results from Table~\ref{tab_bulk} is that
(i) our attempts
to optimize the polar $C2$ phase always ended up in a non-polar structure with
$C2/c$ 
space group; (ii) our PBSsol+$U$ and HSE06 methods give different predictions
regarding which phase is the ground state of bulk Bi$_2$NiMnO$_6$; (iii) there
is a ferromagnetic $R$ phase with large polarization that is competitive with
the phases known so far to exist, and
(iv) no $T$ phase is obtained as a result of our optimizations.
In the following, we provide more details about each of these points.

The original assignment of space group $C2$ to the GS 
phase\cite{Azuma2005JACS} was done after obtaining synchrotron X-ray powder
diffraction peaks that could be indexed as a monoclinic unit cell with 
the lattice parameters quoted in Table~\ref{tab_bulk}.
Because the unit cell was similar to that of BiMnO$_3$, a Rietveld 
refinement was performed by assuming an initial model related to that BiMnO$_3$
structure, and the validity of this model seemed satisfactory.
At the time, the space group BiMnO$_3$ was being described as $C2$, but
later it was shown\cite{Belik2007JACS, Montanari2007PRB, Baettig2007JACS} 
that it is $C2/c$.
While $C2/c$ is centrosymmetric, the $C2$ space group allows for a polarization
along the monoclinic axis; calculations assuming a point-charge
model or using the Berry-phase first-principles theory gave a value of
around 20 $\mu$C/cm$^2$.\cite{Azuma2005JACS,Uratani2006PB,Ciucivara2007PRB,
Zhao2012AIPA, Zhao2012AIPAb}
As far as we know, no experimental measurement of the polarization has been
done for bulk Bi$_2$NiMnO$_6$.

Our attempts to find a metastable $C2$ phase for Bi$_2$NiMnO$_6$ failed.
Every structure we have set with that space group lowered its energy when 
their atoms were allowed to move, ending always in the $C2/c$ structure
displayed in Fig.~\ref{fig_structures}(a).
This also happens for BiMnO$_3$: different $C2$ structures
converge to the
same lower-energy $C2/c$ structure.\cite{Baettig2007JACS, Dieguez2015PRB}
We must however mention that previous first-principles
calculations\cite{Ciucivara2007PRB} did report that a $C2$ phase was found
after optimization (albeit with quite different Wickoff positions than the
experimental work).
We have tried to reproduce those calculations using a methodology similar to
that of Ref.~\onlinecite{Ciucivara2007PRB}, but by allowing enough relaxation
steps
the optimized structure slowly converged to the $C2/c$ one presented
here.
Other first-principles studies of Bi$_2$MnNiO$_6$ were done either
fixing the structure to the experimental one\cite{Uratani2006PB, Zhao2012AIPA}
or to the relaxed first-principles one of Ciucivara and 
coworkers.\cite{Zhao2012AIPAb}
Further support for the $C2/c$ space group is that the reported $C2$ structure
contains two different environments for the Ni$^{2+}$ ions and one for the
Mn$^{+4}$ ions, but there is no
explanation so far for this---there are no signs of charge or orbital
ordering, for example.

To add to the puzzle of the possible paraelectricity of bulk Bi$_2$NiMnO$_6$,
a sizable polarization has indeed been measured in 
Bi$_2$NiMnO$_6$ {\em films}.
The group that synthesized this double perovskite for the first time in
bulk also grew it as a film on SrTiO$_3$ using pulsed laser
deposition.\cite{Sakai2007APL}
They measured a polarization of 5 $\mu$C/cm$^2$ and a magnetic Curie 
temperature of 100 K.
Their films displayed a pseudotetragonal structure with 
$a = b = 3.91$~\AA~(matching the substrate) and $c = 3.87$~\AA, described as
rather different from the bulk one,
while keeping the same rock-salt pattern.\cite{Shimakawa2007JJAP} 
Using a chemical solution deposition method, Lai {\em et al.}\cite{Lai2012IF} 
grew Bi$_2$NiMnO$_6$ films with and without SrTiO$_3$ buffer layers 
on a Pt$(111)$/Ti/SiO$_2$/Si$(100)$ substrate, obtaining polarizations
around 6 and 8 $\mu$C/cm$^2$.
Again, the situation can be compared to that of BiMnO$_3$, where polarizations
between 9 and 23 $\mu$C/cm$^2$ have been reported for 
films,\cite{Son2008APL, DeLuca2013APL, Jeen2011JAP} even if the bulk is 
nonpolar.
For BiMnO$_3$ we proposed that those measurements might be
related to the formation of film phases under strain that are 
polar.\cite{Dieguez2015PRB} 

We move on now to the issue of the predicted ground state by different 
methodologies.
Table~\ref{tab_bulk} shows that the three local minima of the energy surface
of Bi$_2$NiMnO$_6$ lie within only 30 meV per formula unit (five-atom group
of the standard perovskite unit cell;
in our case, BiNi$_{1/2}$Mn$_{1/2}$O$_3$).
Our DFT+$U$ method predicts a higher energy for the
GS phase than for the other two;
this happened too for BiMnO$_3$,\cite{Dieguez2015PRB} where the situation
is corrected by the HSE06 hybrid.
We have shown in the past that different exchange-correlation functionals
predict almost the same minima of the energy surface of 
BiMnO$_3$\cite{Dieguez2015PRB} and BiFeO$_3$,\cite{Dieguez2011PRB} although
how those minima are ordered in energy can vary from functional to functional.
As for BiMnO$_3$, here we use the fast DFT+$U$ method when we are 
interested in finding possible energy minima, or when we are interested
in energy differences between very similar structures, while we will resort
to the more accurate and slow HSE06 when it is important to evaluate energy
differences between different phases.

Both DFT+$U$ and HSE06 predict that there exists a metastable
ferromagnetic rhombohedral structure with a polarization around
70~$\mu$C/cm$^2$ when computed using the Berry-phase 
formalism (we have checked that typical antiferromagnetic alignments
are higher in energy; details are given later for similar films).
This is a structure like that of bulk BiFeO$_3$, but where the superimposed
rock-salt pattern of Mn$^{4+}$ and Ni$^{2+}$ causes a reduction from 
the $R3c$ space group to the $R3$ symmetry; it is represented in 
Fig.~\ref{fig_structures}(c).
According to first principles calculations, a similar structure is also
metastable for BiCoO$_3$\cite{Dieguez2011PRL} and
BiMnO$_3.$\cite{Dieguez2015PRB}

\begin{figure}
\centering
\includegraphics[width=80mm]{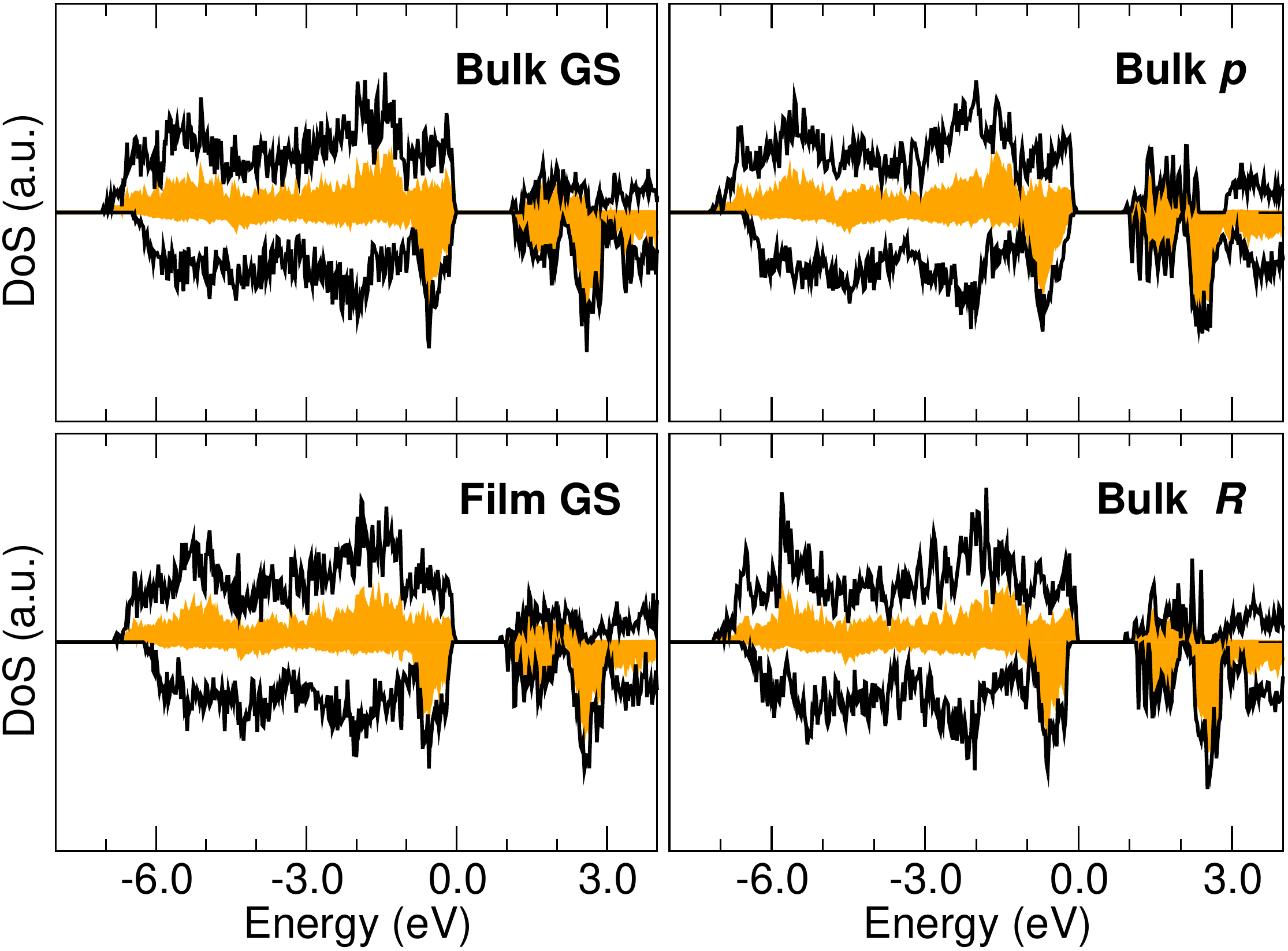}
\caption{(Color online.)
DFT+$U$ density of states for the three bulk phases that are local minima of 
the energy, and for the GS epitaxial film at in-plane lattice parameter of
4~\AA.
The lines represent the total density of states, and the shaded areas 
correspond to its projection onto the $d$ orbitals of Ni and Mn (hybridized
mostly with the $p$ orbitals of O).
}
\label{fig_dos}
\end{figure}

Regarding the electronic structure, the GS, $p$, and $R$ phases display
similar density of states profiles, as shown in Fig.~\ref{fig_dos} (this
was also the case for different phases in BiFeO$_3$\cite{Dieguez2011PRB}).
The band gap in all cases is around 1 eV.

Unlike in BiCoO$_3$, BiFeO$_3$, and BiMnO$_3$, no 
$T$ structures appeared as local minima of the energy 
in our Bi$_2$NiMnO$_3$ search.
We relaxed variations of the $T$ configurations of those other materials
with the added rock-salt pattern of Mn$^{4+}$ and Ni$^{2+}$, but the resulting
structure was always one of the other three local minima. 

\subsection{Epitaxial Films}

One possible way to stabilize a metastable phase of the bulk of a material
is to grow it as a thin film on a substrate; in this way, the epitaxial
misfit strain acts as a 
handle to vary the relative energies of the possible bulk phases.
We have simulated coherent epitaxial (001) films of Bi$_2$NiMnO$_3$ by doing
calculations of the bulk material where we impose mechanical boundary
conditions determined by the lattice constant of the substrate, 
assumed to display in-plane square symmetry (this is indeed 
the case for many perovskite substrates cut perpendicularly to one
of the principal axes).

\begin{figure}
\centering
\includegraphics[width=85mm]{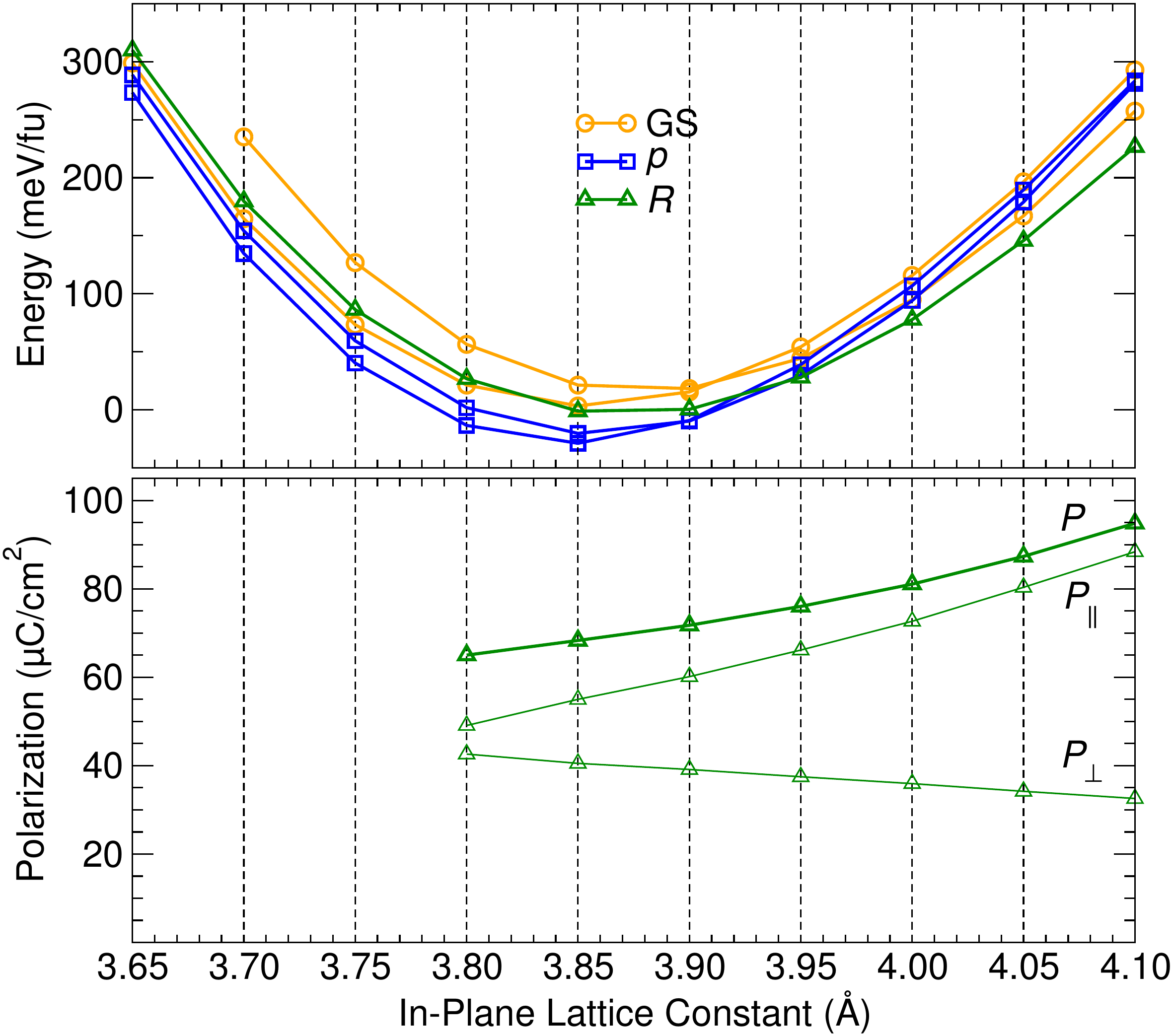}
\caption{(Color online.)
Top: DFT+$U$ energy of relaxed (001) films as a function of the in-plane
lattice parameter;
the films are adaptations of the bulk GS (two possible orientations), $p$
(two possible orientations), and $R$ (one inequivalent orientation)
phases to the mechanical
boundary conditions imposed by the square symmetry of the
substrate.
Bottom: magnitude of the polarization of the $R$ films ($P$), its component
on the film ($P_{||}$), and its component perpendicular to the film 
($P_\perp$).
}
\label{fig_films}
\end{figure}

As a starting point, we adapted the three bulk phases described in the previous
section to the in-plane square symmetry. There are two inequivalent ways to
do this for the GS and $p$ phases, and one way for the $R$ phase, as shown
in Fig.~\ref{fig_structures}.
This causes small distortions to bring the in-plane lattice vectors to form
a $90^{\circ}$ angle and to be of the same magnitude (in the adapted $p$ and
$R$ phases) or of a ratio of magnitudes equal to two (in the GS phase).
In all cases, those distortions cost only a few meV per formula unit.
Then, we do calculations in which we expand or contract the lattice vectors 
to mimic the effect of squared substrates with different lattice constants.
We do this in intervals of 0.05~\AA, and we use the Wickoff positions and
out-of-plane lattice vector of the previous geometry as a starting point of the
next geometry relaxation.
In this way, we arrive at a graphic of the energy of the films as a function
of in-plane lattice constant that we show in Fig.~\ref{fig_films} (top).
These calculations were done using the DFT+$U$ method.

For strains around the minimum of the energy curves the adopted configuration
is paraelectric.
However, 
for high enough tensile strains the
$R$ phase has lower energy than the other phases. 
These strains correspond to in-plane lattice constants of the
order of 4~\AA, so this phase is expected to appear if the films are grown over
perovskite oxides such as BaTiO$_3$ or 
PbZr$_{1-x}$Ti$_{x}$O$_{3}$ (PZT).
Figure~\ref{fig_films} (bottom) shows that the computed polarization of the
films is similar to the 70 $\mu$C/cm$^2$ of the bulk phase.
The electronic structure of the films is very similar to that of the bulk,
as illustrated in Fig~\ref{fig_dos} for the film with in-plane lattice parameter
$a = 4$~\AA.

\begin{figure}
\centering
\includegraphics[width=80mm]{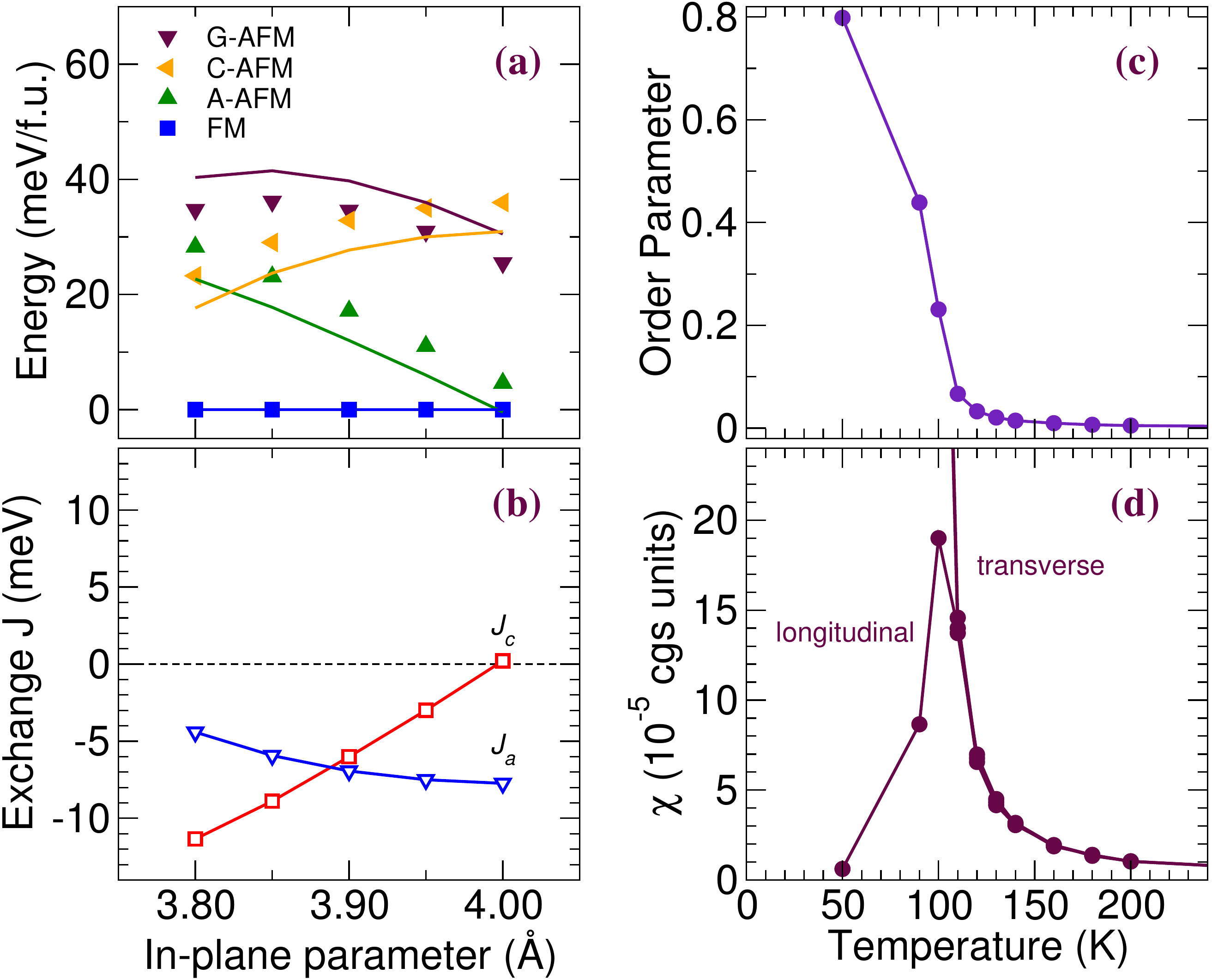}
\caption{(Color online.)
Magnetic properties of $R$ films:
(a) DFT+$U$ values of the energy of different magnetic arrangements (symbols)
and fit to a Heisenberg model (lines); (b) exchange coupling constants $J$
that result from this fit; (c) ferromagnetic order parameter as a function of
temperature obtained from the Heisenberg model when $a = 3.95$~\AA;
and (d) longitudinal
and transverse magnetic
susceptibility obtained from the Heisenberg model when $a = 3.95$~\AA.
}
\label{fig_magnetism}
\end{figure}

All GS, $p$, and $R$ films at tensile epitaxial strains have magnetic 
cations with magnetic moments around $3 \mu_{\rm B}$ for Mn$^{4+}$ and 
around $2 \mu_{\rm B}$ for Ni$^{2+}$.
Figure ~\ref{fig_magnetism}(a) shows that below in-plane lattice parameters
around 4.05~\AA~ferromagnetism prevails over the
alternative 
antiferromagnetic orderings typical of perovskites: G type 
(antiferromagnetism in-plane and out-of-plane), C type (ferromagnetism 
out-of-plane), and A type (ferromagnetism in-plane).
This is not surprising, since the network of transition-metal ions and 
oxygens connecting them has angles and bond lengths similar to those of
the bulk network, known to be ferromagnetic.

Following a prescription we have described in earlier 
articles,\cite{Dieguez2011PRL, Escorihuela2012PRL, Dieguez2015PRB} we have
used the DFT+$U$ energy differences between magnetic arrangements to fit
a simplified Heisenberg model.
This model has as parameters two exchange constants $J_a$ and $J_c$ that
take into account the strength of the magnetic interaction
between neighboring Ni--Mn pairs in plane and
out of plane, respectively; their values are represented in 
\ref{fig_magnetism}(b).
A Monte Carlo method on a lattice of $20 \times 20 \times 20$ spins
was used with this Heisenberg model to study the
behaviour of the magnetic ordering with temperature.
Doing this, we found that the ferromagnetic order parameter takes values
other than zero for temperatures below a Curie point of around 100 K, 
as shown in \ref{fig_magnetism}(c); this is in agreement with experimental
measurements done in Bi$_2$NiMnO$_3$ films.\cite{Sakai2007APL}
The magnetic susceptibility computed from the Monte Carlo simulations with
this Heisenberg model is plotted in Fig.~\ref{fig_magnetism}(d).

\begin{figure}
\centering
\includegraphics[width=85mm]{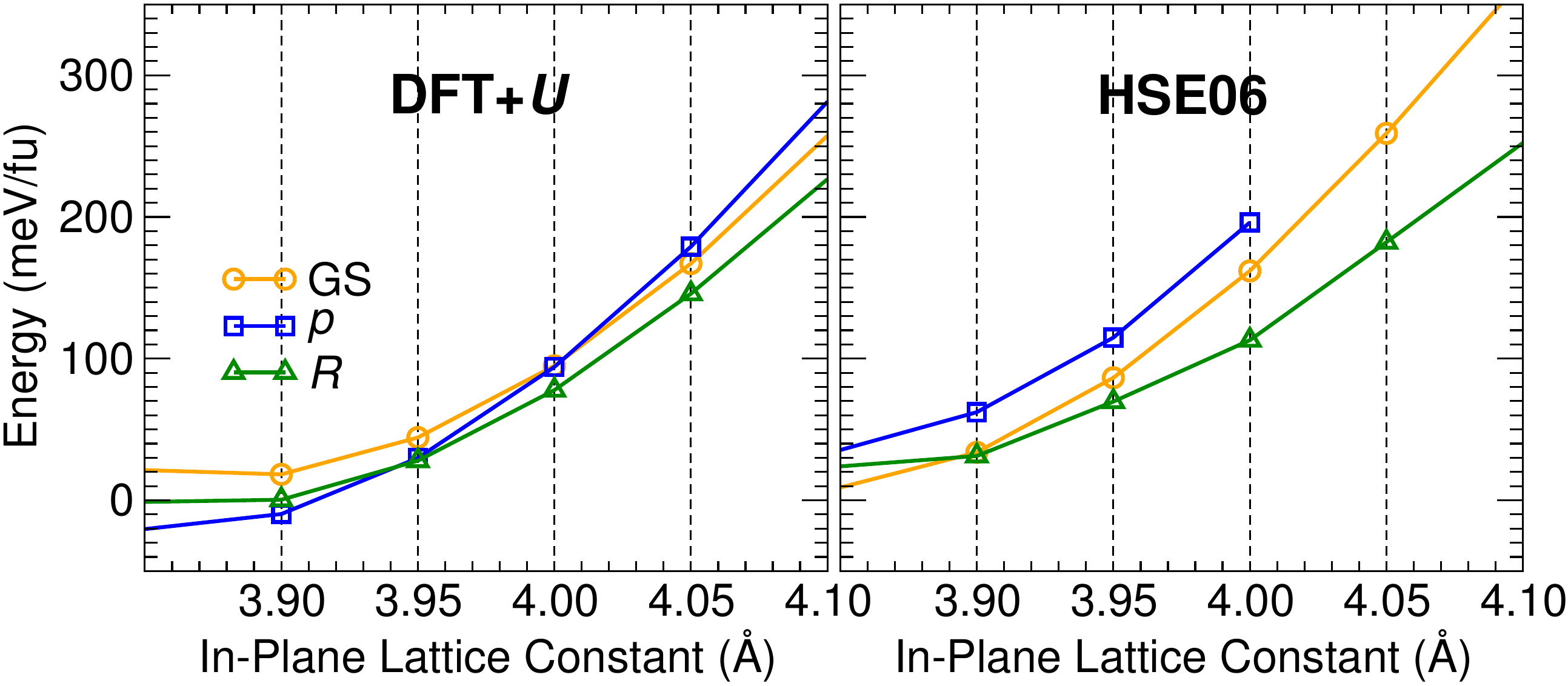}
\caption{(Color online.)
Comparison of the energy of relaxed films as a function of the in-plane
lattice parameter when using DFT+$U$ (left) and HSE06 (right).
When two orientations of the films are possible (GS and $p$ phases) we have
done HSE06 calculations for the one with lowest DFT+$U$ energy
in most of the range of in-plane lattice parameters.
}
\label{fig_filmsHSE06}
\end{figure}

The results for films presented so far were obtained using
DFT+$U$ calculations.
As for BiMnO$_3$,\cite{Dieguez2015PRB} this methodology does not
resolve the close energy differences between phases in agreement with 
experiment, but the HSE06 hybrid functional does.
When we applied it to do computations for 
tensile films of Bi$_2$NiMnO$_6$ it also predicted that
the $R$ phase is the most stable one for large enough strains, as shown
in Fig.~\ref{fig_filmsHSE06}.

\begin{figure}
\centering
\includegraphics[width=85mm]{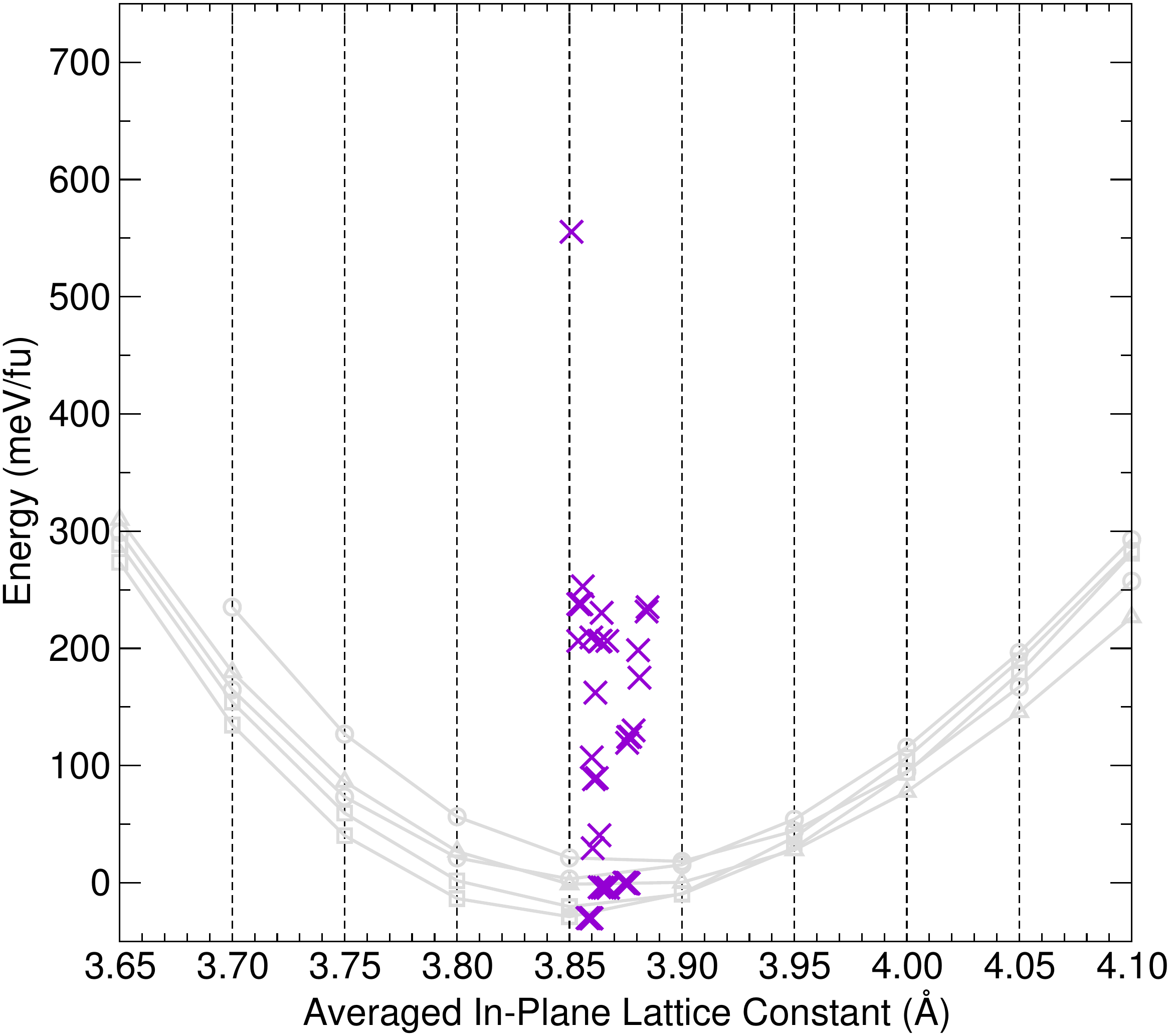}
\caption{(Color online.)
The crosses correspond to the 
energy of the bulk optimized structures that result after performing an
annealing on each of the film structures recorded in Fig.~\ref{fig_films} as
described in the text.
In gray we have copied the data of Fig.~\ref{fig_films} for reference.
}
\label{fig_annealing}
\end{figure}

In order to further explore the energy surface of bulk Bi$_2$NiMnO$_6$ in the
search of minima that might be relevant in films, we did one more set
of calculations.
We took every film structure represented by a point in 
Fig.~\ref{fig_films}, removed the epitaxial constraints, performed
a few steps of molecular dynamics to allow it to explore its surroundings, and
relaxed this structure until the forces and stresses were almost zero. 
During this annealing process the atoms visited structures that were up to a
few eV/fu higher in energy than the ground state.
By the end of the search, most of the initial structures had converged to 
a configuration with forces below 0.015 eV/\AA, and these are 
represented in Fig.~\ref{fig_annealing}, where the energy with respect to 
the bulk ground state is plotted as a function of 
the average of the two closest in-plane lattice
constants of the optimized structure.
We see 
that the low-energy points are near the minima of the film curves of 
Fig.~\ref{fig_films}, showing that
in several cases releasing the epitaxial constraints just takes the system
to one of the
three bulk phases described in this work.
After analyzing the rest of the crosses in Fig.~\ref{fig_annealing}, it turns
out that they also correspond to one of the GS, $p$, or $R$ phases,
but with a different electronic configuration
(e.g.,
the value at around 550 meV/fu is a R structure where all ions show 
zero magnetic moments).
In all, we are confident that no other low energy structures exist for typical
in-plane lattice constants, and in particular in the region where the films
are found to be ferroelectric and ferromagnetic.

\section{Summary and Conclusions}

Our first-principles calculations for bulk Bi$_2$NiMnO$_6$ are consistent
with a non-polar crystal structure of space group $C2/c$.
Previous reports pointed out to a $C2$ polar space group, but the reasons 
stated in Section III.A lead us to believe that, as it happened with BiMnO$_3$,
this is not correct.

Our calculations also show that when Bi$_2$NiMnO$_6$ is grown on a
(001)-oriented perovskite substrate of materials such as BaTiO$_3$ or PZT
the epitaxial 
strain should favour a phase that is both ferroelectric and ferromagnetic.
The polarization of these films is around 70~$\mu$C/cm$^2$, similar to that
of the most used ferroelectric materials. 
The films are predicted to be ferromagnetic with magnetic moments
of 2.5~$\mu_{\rm B}$ per formula unit
and a Curie temperature of around 100 K.
Thus, our simulations predict that, in thin film form, Bi$_2$NiMnO$_6$ is one
of the very few known magnetoelectric multiferroics with a strong
ferromagnetic order.


\section*{Acknowledgements}
{O.D. acknowledges funding from the Israel Science Foundation through
Grants 1814/14 and 2143/14.
J.\'I. was financially supported by the Luxembourg National Research Fund
through the Pearl (Grant No. P12/485315) and Core (Grant No. C15/MS/10458889)
programs.
}



\end{document}